\documentclass[12pt,preprint]{aastex}

\def\sun{\odot}

\begin{document}
\title{The Ratio of Retrograde to Prograde Orbits: A Test for Kuiper
  Belt Binary Formation Theories} \shortauthors{Schlichting & Sari}
  \shorttitle{Prograde and Retrograde Orbits of Kuiper Belt Binaries}
  \author{Hilke E. Schlichting\altaffilmark{1} and Re'em Sari\altaffilmark{1,2} }
\altaffiltext{1}{California Institute of Technology, MC 130-33, Pasadena, CA
  91125} 
\altaffiltext{2}{Racah Institute of Physics, Hebrew University,
  Jerusalem 91904, Israel}
\email{hes@astro.caltech.edu, sari@tapir.caltech.edu}

\begin{abstract} 
With the discovery of Kuiper Belt binaries that have wide separations and
roughly equal masses new theories were proposed to explain their
formation. Two formation scenarios were suggested by Goldreich and
collaborators: In the first, dynamical friction that is generated by a sea
of small bodies enables a transient binary to become bound ($L^2s$ mechanism);
in the second, a transient binary gets bound by an encounter with a third body
($L^3$ mechanism).

We show that these different binary formation scenarios leave their own unique
signatures in the relative abundance of prograde to retrograde binary
orbits. This signature is due to stable retrograde orbits that exist much
further out in the Hill sphere than prograde orbits. It provides an excellent
opportunity to distinguish between the different binary formation scenarios
observationally.

We predict that if binary formation proceeded while sub-Hill velocities
prevailed, the vast majority of all comparable mass ratio binaries have
retrograde orbits. This dominance of retrograde binary orbits is a result of
binary formation via the $L^2s$ mechanism, or any other mechanism that
dissipates energy in a smooth and gradual manner. For super-Hill velocities
binary formation proceeds via the $L^3$ mechanism which produces a roughly
equal number of prograde and retrograde binaries. These predictions assume
that subsequent orbital evolution due to dynamical friction and
dynamical stirring of the Kuiper belt did not alter the sense of the binary
orbit after formation.
\end{abstract}


\section{INTRODUCTION}
The detection of comparable mass binaries with wide separations in the Kuiper
Belt called for new theories explaining their formation
\citep[e.g.][]{W02,GLS02,F04,A05,A07}. Their existence cannot be explained
with a formation scenario involving a collision and tidal evolution, as has
been proposed for the formation of the Moon and Charon \citep{HD75,CW76,MK89},
since it cannot provide the current angular momentum of the binary system. In
a formation scenario proposed by \citet{W02} two Kuiper Belt objects (KBOs)
collide with each other inside the Hill sphere of a third. However, in the
Kuiper Belt, gravitational scattering between the two intruders is about 100
times\footnote{For this estimate we used $\alpha\sim 10^{-4}$ and assumed that
the velocity dispersion of the KBOs at the time of binary formation is less
than their Hill velocity, see \S 2 for details} more common than a
collision. Therefore, three body gravitational deflection ($L^3$ mechanism),
as proposed by \citet{GLS02}, should dominate the binary formation over such
collisional scenario.  A second binary formation scenario that has been
suggested by \citet{GLS02} consists of the formation of a transient binary
that gets bound with the aid of dynamical friction from a sea of small
bodies. We call this the $L^2s$ mechanism.  In the formation scenario of
\cite{A05} and \cite{A07} the existence of long lived transient binaries that
spend a long time in their mutual Hill sphere, near a periodic orbit, is
responsible for the creation of Kuiper Belt binaries (KBBs). Finally,
\citet{F04} proposed a binary formation mechanism that involves a collision
between two large KBOs. This collision creates a small moon that is replaced
in an exchange reaction by a massive body with high eccentricity and large
semi-major axis.

In this paper, we show that the $L^2s$ and $L^3$ mechanism leave unique
signatures in the relative abundance of prograde to retrograde binary
orbits. The $L^2s$ mechanism dominates over the $L^3$ mechanism for sub-Hill
velocities \citep{SR07}. We argue that binaries that form from dynamically
cold KBOs by the $L^2s$ mechanism have retrograde orbits. This is due to the
existence of stable retrograde binary orbits with modified Jacobi constants
similar to that of unbound KBOs on circular orbits that have impact parameters
that correspond to distances of closest approach of less than the Hill
radius. No equivalent prograde orbits exist
\citep[e.g.][]{H70,I79,ZI88,HB91,HK97}. Since dynamical friction only
gradually increases the modified Jacobi constant (for a binary this
corresponds to gradually increasing the absolute value of the binding energy),
all binaries that form via the $L^2s$ mechanism, or any other mechanism that
dissipates energy in a smooth and gradual manner, will start with modified
Jacobi constants close to that of unbound KBOs that penetrate the Hill sphere
and hence have retrograde orbits.  For super-Hill KBO velocities, only the
$L^3$ mechanism can form tight binaries that tend to survive \citep{SR07}. The
fact that retrograde orbits are stable for larger semi-major axes is no longer
of importance since only tight binaries are saved from break up. This,
therefore, leads to the formation of a roughly equal number of prograde and
retrograde binaries for super-Hill KBO velocities.

Our paper is structured as follows: In \S 2 we outline our assumptions,
explain our choice of parameters and define variables that will be used
throughout this paper. We calculate the ratio of prograde to retrograde binary
orbits for the $L^2s$ and $L^3$ mechanism and predict the relative abundance
of prograde to retrograde orbits for sub-Hill and super-Hill KBO velocities in
\S 3. We compare our predictions with observations in \S 4. Discussion and
conclusions follow in \S 5.

\section{DEFINITIONS AND ASSUMPTIONS}
The Hill radius denotes the distance from a body at which the tidal forces due
to the Sun and the gravitational force due to the body, both acting on a test
particle, are in equilibrium. It is given by
\begin{equation}\label{e1}
R_{H} \equiv a_{\sun} \left( \frac{m_1+m_2}{3 M_{\sun}}\right) ^{1/3} 
\end{equation}
where $m_1$ and $m_2$ are the masses of the two KBOs, $a_{\sun}$ is their
semi-major axis around the Sun and $M_{\sun}$ the mass of the Sun. Our
definition of the Hill radius differs from that used by \citet{SR07} since we
include the combined mass of both KBOs here. We chose to do so since it will
make comparisons with works by other authors easier.

We use the `two-group approximation' \citep{GLS02,GLS04} which consists of the
identification of two groups of objects, small ones, that contain most of the
total mass with surface mass density $\sigma$, and large ones, that contain
only a small fraction of the total mass with surface mass density $\Sigma \ll
\sigma $. We assume $\sigma \sim 0.3 \rm{g~cm^{-2}}$ which is the
extrapolation of the minimum-mass solar nebular \citep{H81} to a heliocentric
distance of $40\rm{AU}$. Estimates from Kuiper Belt surveys
\citep{TJL01,TB03,PK08,F08,FH08} yield $\Sigma \sim 3 \times 10^{-4}
\rm{g~cm^{-2}}$ for KBOs with radii of $R \sim 100~\rm{km}$. We use this value
of $\Sigma$, assuming that $\Sigma$ during the formation of KBBs was the same
as it is now. Our choice for $\Sigma$ and $\sigma$ is also consistent with
results from numerical coagulation simulations by \citet{KL99}.

Large bodies grow by the accretion of small bodies. Large KBOs viscously stir
the small bodies, increasing the small bodies' velocity dispersion $u$. As a
result $u$ grows on the same timescale as $R$ provided that mutual collisions
among the small bodies are not yet important.  In this case, $u$ is given by
\begin{equation}\label{e2}
\frac{u}{v_H} \sim \left( \frac{\Sigma}{\sigma \alpha} \right)^{1/2} \sim 3
\end{equation}
where $\alpha = R/R_{H}\sim 10^{-4}$ at $40\rm{AU}$ \citep{GLS02}. $v_H$ is
the Hill velocity of the large bodies which is given by $v_H = \Omega R_H$
where $\Omega$ is the orbital frequency around the sun. The velocity $v$ of
large KBOs increases due to mutual viscous stirring, but is damped by
dynamical friction from a sea of small bodies such that $v < u$. Balancing
the stirring and damping rates of $v$ and substituting for $u$ from equation
(\ref{e2}), we find
\begin{equation}\label{e3}
\frac{v}{v_H} \sim \alpha^{-2} \left(\frac{\Sigma}{\sigma}\right)^{3} \sim
0.1.
\end{equation} 
For our choice of parameters, we have sub-Hill KBO velocities during the epoch
of formation of bodies with $R\sim 100\rm{km}$. We therefore focus our work on
the shear-dominated velocity regime ($v \ll v_{H}$). However, we discuss how
our results would be modified if $v \gg v_H$.

\section{PROGRADE VERSUS RETROGRADE BINARY ORBITS}

\subsection{Sub-Hill velocities: $v \ll v_H$}
\begin{figure}[htp]
\centerline{
\includegraphics[ scale=1.6]{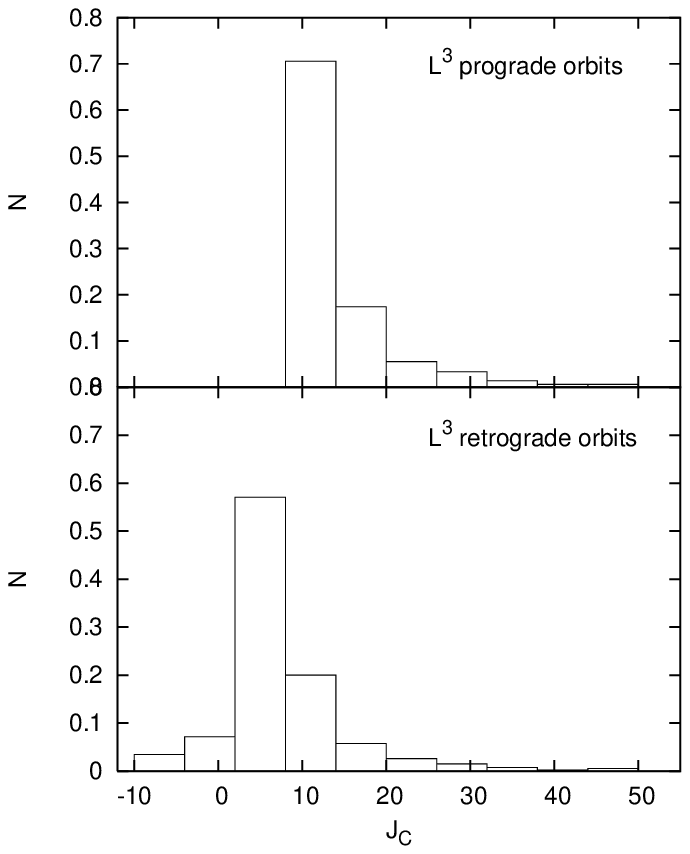}
}
\caption{Histogram of modified Jacobi constants, $J_C$, of prograde and
  retrograde KBBs that formed via three body gravitational deflection, $L^3$
  mechanism, for $v \ll v_H$. Each histogram is normalized to unity, but
  overall retrograde orbits are twice as abundant as prograde orbits. Note,
  prograde binaries exist only for $J_C \gtrsim 9$ whereas retrograde binaries
  exist also for $J_C \lesssim 9$.}
\label{fig1} 
\end{figure}

The disk of KBOs is effectively two-dimensional in the shear-dominated
velocity regime ($v \ll v_H$), since the growth of inclinations is suppressed
\citep{WS1993,R2003,GLS04}. We therefore restrict our calculations for the
shear-dominated velocity regime to two dimensions. Since we are interested in
close encounters among the KBOs, their interaction is well described by Hill's
equations \citep{H78,GT80,HP86}. In Hill coordinates the equations of motion
of the two KBOs can be decomposed into their center of mass motion and their
relative motion with respect to one another. The modified Jacobi constant is
exactly conserved in the Hill formalism, but the Hill formalism itself is an
approximation to the general three body problem. It assumes that the masses of
body 1 and 2 (in our case the two KBOs) are much less than that of the Sun. We
use the standard Hill coordinate system and reference frame as in \citet{HP86}
and \citet{Ida90}. In this rotating frame the direction of the $x$-axis is
given by the line connecting the Sun and the center of mass of the two KBOs
such that the positive $x$ direction is pointing away from the Sun. The
$y$-axis is perpendicular to the $x$-axis pointing in the direction of motion
of the KBOs' center of mass around the Sun. In Hill coordinates the modified
Jacobi constant is
\begin{equation}\label{e4}
J_C=3x^2 + \frac{6}{(x^2+y^2)^{1/2}} -\dot{x}^2 -\dot{y}^2
\end{equation}  
where $x$ and $y$ correspond to the relative separation between the two KBOs
in the $x$ and $y$ direction respectively \citep{HP86}. Length has been scaled
by $R_{H}$ and time by $\Omega^{-1}$. In Hill coordinates the Lagrangian
points $L_1$ and $L_2$ are located at $(-1,0)$ and $(+1,0)$ respectively,
where we define $L_1$ as the Lagrangian point located between the KBO and the
Sun. Their modified Jacobi constants are $J_C(L_1)=J_C(L_2)=9$. From equation
\ref{e4} we can see that tight binaries with small separations have $J_C \gg
9$. We call a binary orbit prograde if its angular momentum about the binary
center of mass, as viewed in the non rotating frame, is in the same direction
as the orbital angular momentum of the binary around the Sun. If the binary
angular momentum is in the opposite direction to the orbital angular momentum
of the binary around the Sun, the orbit is called retrograde. Several authors
recognized that planar retrograde orbits are stable for larger semi-major axes
than prograde orbits \citep[e.g.][]{H70,I79,ZI88,HB91,HK97}.  A prograde
binary with an initially circular orbit becomes unbound for $a \gtrsim 0.49
R_H$ where $a$ is the initial semi-major axis of the mutual binary orbit
\citep{HB91}. This implies that prograde orbits with modified Jacobi constants
less than that of the Lagrangian points $L_1$ and $L_2$ are unbound. In
contrast to the prograde case, there exist stable retrograde binary orbits
with $J_C \lesssim J_C(L_1)=J_C(L_2)=9$. This result is also shown in Figures
\ref{fig1}. Figure \ref{fig1} shows histograms of $J_C$ for prograde and
retrograde binaries that formed by $L^3$ mechanism from KBOs with initially
circular orbits around the Sun. In the reminder of this paper we discuss the
stability of prograde and retrograde orbits in terms of $J_C$ and not
semi-major axis since the latter is not well defined (i.e. it is not a
constant of motion) for wide orbits with $a \sim R_H$. The modified Jacobi
constant for two KBOs that approach each other from infinity is
\begin{equation}\label{e14}
J_C=3x^2 -\dot{x}^2 -\dot{y}^2=\frac{3}{4}b^2 - e^2
\end{equation} 
where $b$ is the initial separation between the two KBOs in the $x$ direction
and $e$ is the relative eccentricity in Hill units given by $\vert
\mathbf{e_1}-\mathbf{e_2} \vert$ where $\mathbf{e_1}$ and
$\mathbf{e_2}$ are the eccentricity vectors of body 1 and body 2
respectively. Only KBOs with $b$ ranging from $1.7 R_H$ to $2.5 R_H$ penetrate
each others Hill sphere if started on circular orbits. From equation
(\ref{e14}) we have therefore that only KBOs with $2.2 \le J_C \le 4.7$ have a
distance of closest approach of $R_H$ or less provided that they started on
circular orbits around the Sun.

\subsubsection{$L^2s$ Mechanism}
In the $L^2s$ mechanism KBBs form from transient binaries that become bound
with the aid of dynamical friction from a sea of small bodies. This dynamical
friction provides a gentle force that damps the random velocity of large
KBOs. For typical parameters, the dynamical friction force only extracts a
small fraction of energy over an orbital timescale. Therefore, KBBs that form
via the $L^2s$ mechanism, or any other mechanism that dissipates energy
gradually, have initially modified Jacobi constants similar to that of the
unbound KBOs that penetrate within the Hill sphere. As mentioned above, for
KBOs that started on circular orbits around the Sun this corresponds to $2.2
\le J_C \le 4.7$. However, only stable retrograde orbits exist with $J_C
\lesssim 9$. This implies that all KBB that form this way must have retrograde
orbits since no stable prograde orbits exist for $J_C \lesssim 9$.  Once a
binary is formed, dynamical friction increases the modified Jacobi constant and
the absolute value of the binary binding energy. We confirm that all binaries
that form from KBOs on initially circular orbits around the Sun via the
$L^2_s$ mechanism are retrograde by numerical integrations that are presented
below.

Since it is not feasible to examine the interactions with each small body
individually, their net effect is modeled by an averaged force which acts to
damp the large KBOs' non-circular velocity around the Sun. We parameterize the
strength of the damping by a dimensionless quantity $D$ defined as the
fractional decrease in non-circular velocity due to dynamical friction over a
time $\Omega^{-1}$:
\begin{equation}\label{e5}
D\sim \frac{ \sigma}{\rho R} \left(\frac{u}{v_H}\right)^{-4} \alpha^{-2} \sim
\frac{\Sigma}{\rho R} \alpha^{-2} \left(\frac{v}{v_H}\right)^{-1}.
\end{equation} 
The first expression is simply an estimate of dynamical friction by a sea of
small bodies assuming $u>v_H$. The second expression describes the mutual
excitation among the large KBOs for $v \ll v_H$. $v$ achieves a quasi-steady
state on a timescale shorter than at which $R$ grows since only a subset of
the deflected bodies are accreted. The stirring among the large KBOs can
therefore be equated to the damping due to dynamical friction (for a detailed
derivation see \citet{GLS04}).

Since the growth of inclinations is suppressed in the shear-dominated velocity
regime the disk of KBOs is effectively two-dimensional
\citep{WS1993,R2003,GLS04}. We therefore restrict this calculation to two
dimensions. In Hill coordinates the relative motion of two equal mass KBOs,
including the dynamical friction term, is governed by
\begin{equation}\label{e6} 
\ddot{x}-2 \dot{y}
-3x=-\frac{3 x}{(x^2+y^2)^{3/2}}-D \dot{x}
\end{equation} 
\begin{equation}\label{e7}\ddot{y}+2
\dot{x}=-\frac{3 y}{(x^2+y^2)^{3/2}} -D (\dot{y}+1.5x).
\end{equation} 
Length has been scaled by $R_{H}$ and time by $\Omega^{-1}$. Equations
(\ref{e6}) and (\ref{e7}) are integrated for different values of $D$ and
impact parameters ranging from $1.7 R_H$ to $2.5 R_H$ with equal step
size. Impact parameters outside this range result in a distance of closest
approach between the two KBOs of more than $R_H$.

\begin{figure} [htp]
\centerline{\includegraphics[ scale=1.2]{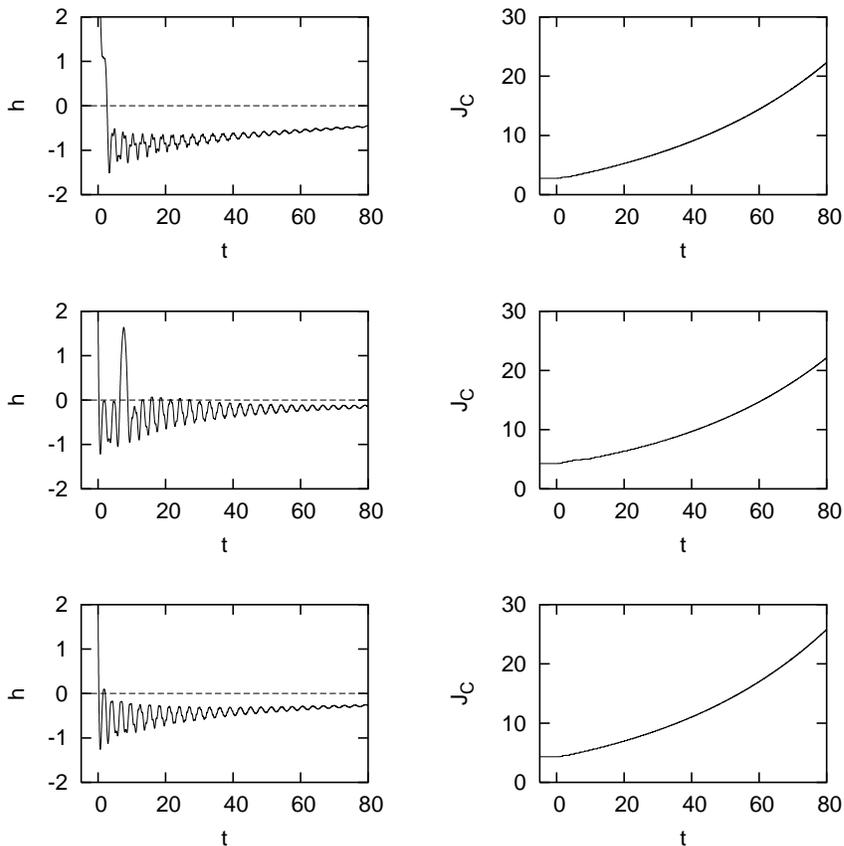}}
\caption{Three examples of KBO encounters in the $L^2s$ mechanism for $v \ll
  v_H$ and $D=0.01$ that result in the formation of a binary. The plots on the
  left and right hand side show the evolution of the specific angular
  momentum, $h$, and modified Jacobi constant, $J_C$, as a function of time
  respectively. The time $t=0$ corresponds to the time at which $y=0$ if the
  relative KBO velocity is solely due to the Keplerian sheer (i.e. ignoring
  the actual gravitational interaction between the bodies). The evolution of
  $h$ and $J_C$ is shown until the binary separation has decreased to $0.1
  R_H$ or less. These examples show that the sense of rotation is practically
  preserved. $h$ displays large variations right after capture caused by solar
  tides. The most extreme case of angular momentum sign change for bodies that
  form binaries found in our simulations is displayed in the second of the
  three examples. The angular momenta of the binaries are all negative
  corresponding to retrograde binary orbits. In fact all binaries that form
  via the $L^2s$ mechanism in our numerical integrations display retrograde
  orbits. Dynamical friction shrinks the binary separation leading to a
  decrease in the magnitude of $h$ and an increase of $J_C$ with time. The
  modified Jacobi constant of the newly formed binaries is smaller than
  $J_C(L1)=9$ which explains why all their orbits are retrograde (see \S
  3.1.1. for details).}
\label{fig2} 
\end{figure}

\begin{figure} [htp]
\centerline{\includegraphics[ scale=1.2]{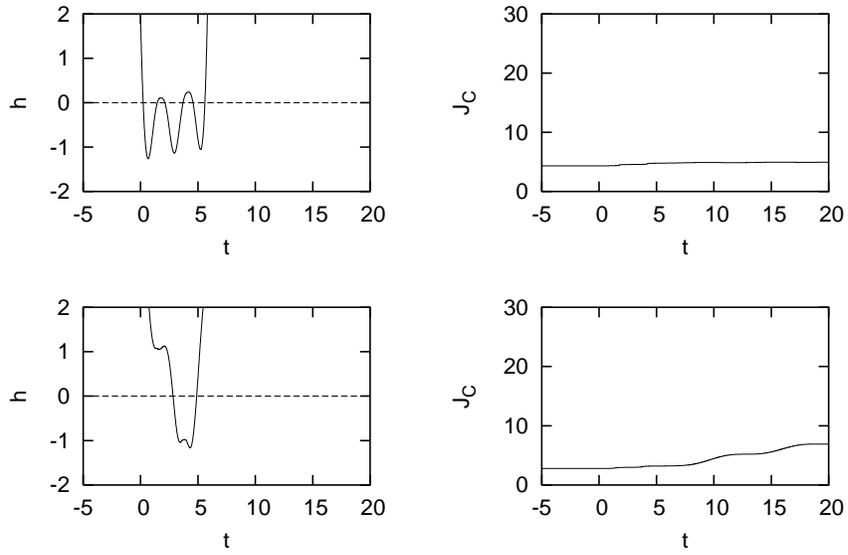}}
\caption{Same as in Figure \ref{fig3} but for two examples of KBO encounters
  in the $L^2s$ mechanism for $v \ll v_H$ and $D=0.01$ that do not result in
  the formation of a binary. As a result of the Keplerian sheer, KBOs
  encounter and leave each other with positive $h$.}
\label{fig3} 
\end{figure}

For $D=0.01$, we performed 20000 integrations. About two percent of these
integrations resulted in the formation of a binary. Figure \ref{fig2} shows
three examples of the evolution of the specific angular momentum and $J_C$ of
binary formation events from our integrations for $D=0.01$. In addition, we
performed integrations for values of $D$ ranging from 0.1 to 0.0004 and find
that, just like in the $D=0.01$ case, only retrograde binaries form. We define
$h$ as the specific angular momentum of the binary in the non rotating
frame. It can be written as $h=x\dot{y}-y\dot{x}+x^2+y^2$ and is related to
the total binary orbital angular momentum, $L$, by $h=(1/m_1+1/m_2) L$. The
time $t=0$ corresponds to the time at which $y=0$ if the relative KBO velocity
is solely due to the Keplerian sheer (i.e. ignoring the actual gravitational
interaction between the bodies). The evolution of $h$ and $J_C$ is shown until
the binary separation has decreased to $0.1 R_H$ or less. Binaries with
separations of $0.1 R_H$ or less are sufficiently tight that perturbations
from the Sun are too weak to flip the sign of the angular momentum. As
expected from our discussion above, the angular momenta of the binaries are
negative corresponding to retrograde binary orbits. In fact all binaries that
form via the $L^2s$ mechanism in our numerical integrations display retrograde
orbits. Dynamical friction shrinks the binary separation. As a result, the
magnitude of the binary angular momenta decreases with time. The right hand
side of Figure \ref{fig2} shows the evolution of the modified Jacobi
constant. Newly formed binaries initially have a modified Jacobi constant $<9$
which is possible only for retrograde binaries. Dynamical friction shrinks the
semi-major axes of the binaries which leads to an increase of $J_C$ with time
while keeping the sense of rotation, i.e. the sign of $h$, fixed. Eventually
the modified Jacobi constant grows to values above $J_C(L1)=J_C(L2)= 9$. For
$J_C\gtrsim 9$ prograde orbits can exist; however all binaries that formed
with the aid of dynamical friction started out with $J_C < 9$ for which only
retrograde orbits are stable. Therefore, all KBB that form via the $L^2s$
mechanism, or any other mechanism that gradually removes energy from transient
binaries, orbit each other in the retrograde sense since otherwise they would
not be able to form in the first place. Figure \ref{fig3} shows the evolution
of $h$ and $J_C$ as a function of time for KBO encounters that did not lead to
the formation of a binary. These examples show that KBOs encounter each other
and leave each other with positive angular momenta. This is a result of the
Keplerian sheer and follows from the definition of $h$.

We have assumed here that all KBOs are initially on circular orbits around the
Sun and have shown that this leads to the formation of exclusively retrograde
binaries in the $L^2s$ mechanism. If, however, the velocity dispersion of the
KBOs is sufficiently large, such that $e$ is of the order of the Hill
eccentricity, bigger impact parameters allow the KBOs to penetrate each others
Hill sphere. In this case, there now exist KBOs that have an initial $J_C$
just a little below $9$ (see equation \ref{e14}) in which case only a small
change in $J_C$ is sufficient for the formation of retrograde and prograde
binaries. Therfore, prograde binaries can form with the aid of dynamical
friction provided that the velocity dispersion of the KBOs is about $v_H$.

Our prediction for the sense of the binary orbit relies on the assumption that
dynamical friction does not alter the sense of the binary orbit in the
subsequent binary evolution. Although we have shown in our simulations that for
our dynamical friction model this is indeed the case, it might be that the
actual behavior of dynamical friction differs from the model implemented here.

\subsubsection{$L^3$ Mechanism}
A transient binary forms when two large KBOs penetrate each other's Hill
sphere. This transient binary must lose energy in order to become
gravitationally bound. In the $L^3$ mechanism the excess energy is carried
away by an encounter with a third massive body. This encounter can provide a
significant change in energy which corresponds to a considerable change in
$J_C$. The modified Jacobi constants of KBBs that form via the $L^3$
mechanism are therefore not constraint to values similar to that of their
initial $J_C$; their orbits can therefore be both prograde and retrograde. We
show that this is indeed the case with numerical integrations discussed below
and determine the ratio of prograde to retrograde orbits for binary formation
via the $L^3$ mechanism.

Our calculation is performed in the shear-dominated velocity regime in two
dimensions. As initial condition, we assume that all bodies are on circular
orbits. We modify Hill's equations \citep{H78,GT80,PH86} to include three
equal mass bodies besides the Sun. The equations of motion, with length scaled
by $R_{H}$ and time by $\Omega^{-1}$, for body 1 are given by
\begin{equation}\label{e8}
\ddot{x}_{1}-2
\dot{y}_{1} -3x_{1}=-\frac{3(x_{1}-x_{2})}
{2((x_{1}-x_{2})^2+(y_{1}-y_{2})^2)^{3/2}}-\frac{3(x_{1}-x_{3})}
{2((x_{1}-x_{3})^2+(y_{1}-y_{3})^2)^{3/2}} \end{equation}
\begin{equation}\label{e9} 
\ddot{y}_{1}+2\dot{x}_{1}=-\frac{3
(y_{1}-y_2)} {2((x_{1}-x_2)^2+(y_{1}-y_2)^2)^{3/2}}-\frac{3(y_{1}-y_{3})}
{2((x_{1}-x_{3})^2+(y_{1}-y_{3})^2)^{3/2}}.
\end{equation}
The subscripts 1, 2 and 3 label the $x$- and $y$-coordinates of KBO 1, 2, and
3 respectively. Similar equations of motion can be obtained for bodies 2 and
3. Resulting binary orbits are calculated by numerically integrating the
equations of motion. We refer the reader to \citet{SR07}
for the exact details of these calculations. 

Figure \ref{fig1} shows histograms of the modified Jacobi constant of prograde
and retrograde binaries that formed via the $L^3$ mechanism. Both histograms
are normalized to unity. As discussed above, we indeed find that prograde
orbits only exist for $J_C \gtrsim 9$. The stability of retrograde orbits
extends below $J_C=9$ down to $J_C \sim -10$. It therefore includes the range
$2.2< J_C <4.7$ with orbits that penetrate the Hill sphere from circular
heliocentric orbits. Unlike the $L^2s$ mechanism, the $L^3$ mechanism does
produce retrograde {\em and} prograde binaries for $v \ll v_H$. We find that
65\% of all binary orbits are retrograde and 35\% prograde (see Figure
\ref{fig4}). Here, we only considered binary formation from three equal mass
bodies that started on initially circular orbits around the Sun. We therefore
caution, that the ratio of prograde to retrograde orbits due to the $L^3$
mechanism might differ for other mass ratios and velocity dispersions.

\subsubsection{The Ratio of Retrograde to Prograde Orbits}
\citet{SR07} have shown that for sub-Hill KBO velocities the ratio of the
$L^3$ to $L^2s$ binary formation rate is
\begin{equation}\label{e10}
\frac{FR_{L^3}}{FR_{L^2s}}=0.05\frac{v}{v_H}.
\end{equation}
Therefore, for sub-Hill KBO velocities, binaries in the Kuiper Belt form
 primarily due to dynamical friction. For our estimate of $(v/v_H)\sim 0.1$,
 we have that $ FR_{L^3}/FR_{L^2s} \sim 0.005 $, in which case $\sim 0.5 \%$
 of all binaries form directly by the $L^3$ mechanism. Since prograde binaries
 can only form via the $L^3$ mechanism, they make up a negligible fraction of
 the total binaries. Below we discuss how a somewhat larger fraction of
 prograde binaries can arise due to exchange reactions with unbound KBOs.

Once a binary is formed its semi-major axis shirks due to dynamical friction
provided by a sea of small bodies. Dynamical friction decreases the orbit of
a KBB that has an orbital velocity $v_B$ at a rate
\begin{equation}\label{e20}
\mathcal{R}_{shrink} \sim D \Omega \sim \frac{\Sigma}{\rho R} \alpha^{-2} \left(\frac{v}{v_H}\right)^{-1}
\end{equation} 
where we assume that $v_B<u$. Exchange reactions or binary break up by passing
KBOs occurs at a rate given by
\begin{equation}\label{e12}
\mathcal{R}_{exchange} \sim \frac{\Sigma }{\rho R} \alpha^{-2} \Omega
\left(\frac{v_B}{v_H}\right)^{-1}.
\end{equation}
The ratio of these two rates is given by
\begin{equation}\label{e13}
\frac{\mathcal{R}_{shrink}}{\mathcal{R}_{exchange}}\sim
\left(\frac{v_B}{v}\right)
\end{equation}
where $v \ll v_H$ and $v_B \gtrsim v_H$. Break up or exchange reactions are
most likely for wide binaries, in which case $v_B \sim v_H$ since $v_B$
increases as the semi-major axis of the mutual binary orbit
decreases. Therefore we have from equation (\ref{e13}) that
$\mathcal{R}_{shrink}/\mathcal{R}_{exchange} \sim (v_H/v) \sim 10$ for our
estimate of $(v/v_H)\sim 0.1$. This implies that $\sim$ 10\% of all binaries
that formed will suffer an exchange reaction or break up. We performed
numerical integrations of binary break up and exchange reactions to obtain a
more accurate estimate and find that only about 3\% of the binaries suffer an
exchange reaction and/or break up. Our order of magnitude calculation,
therefore, slightly over estimates the number of binaries that experience an
exchange reaction and/or break up. Moreover, only a fraction of the these
binaries will end up as binaries with prograde orbits. In conclusion, we
predict that the vast majority ($\gtrsim 97$\%) of comparable mass ratio
binaries will have retrograde orbits if KBO velocities of $v \lesssim 0.1 v_H$
prevailed during binary formation. This prediction assumes that subsequent
orbital evolution due to dynamical friction did not alter the sense of the
binary orbit after formation.

\subsection{Super-Hill Velocity: $v \gg v_H$}
There is some uncertainty in what the actual values of $\sigma$ and $\Sigma$
were during binary formation. For a few times larger value of $ \Sigma $ with
$ \sigma $ unchanged, we enter the regime in which $v$ exceeds the Hill
velocity (this can be seen from equation (\ref{e3})). We discuss here briefly
how this would affect the ratio of prograde to retrograde binary orbits.

\citet{SR07} have shown that, for $v \gg v_H$, only binaries that form with a
binary separation of $R_{crit}=R_H(v_H/v)^2$ or less tend to be saved from
break up. The $L^2s$ mechanism fails in creating binaries with separations
$\sim R_{crit}$ or less since dynamical friction is not able to dissipate
sufficient energy for tight binaries to form. Therefore, the $L^2s$ mechanism
is not important if KBOs have super-Hill velocities. Tight binaries (with
separations less than $\lesssim R_{crit}$), can form via the $L^3$
mechanism. However in this case, the binary formation cross section is
significantly reduced with respect to the sub-Hill velocity regime (see
\citet{N07} and \citet{SR07} for details). The fact that retrograde orbits are
stable for larger semi-major axes is no longer of importance since only tight
binaries tend to survive. We therefore predict that a roughly equal number of
prograde and retrograde binaries form if super Hill velocities prevail. This
prediction is supported by Figure \ref{fig4}. Figure \ref{fig4} shows the
ratio of retrograde binaries with a modified Jacobi constant of $J_C^{min}$ or
larger to the total number of binaries that formed via the $L^3$ mechanism for
$v \ll v_H$. When all binaries are included we find that about $2/3$ have
retrograde orbits. More retrograde than prograde binaries form because
retrograde binary orbits are stable further out in the Hill sphere than
prograde ones. As $J_C^{min}$ increases the fraction of retrograde binaries
decreases reaching a minimum of about $1/3$ for $J_C^{min} \sim9$. This may be
due to the Keplerian sheer which increases the duration of a prograde
encounter between unbound KBOs compared to a retrograde encounter.  The
fraction of prograde and retrograde binaries becomes comparable for $J_C^{min}
\gg 9$ because for such binaries neither the Keplerian sheer nor the increased
stability of retrograde orbits are important. This is the relevant regime for
binaries that form for $v \gg v_H$ since these large modified Jacobi constants
correspond to tight binaries, which are the only binaries that are saved from
break up if super-Hill velocities prevail.

\section{COMPARISON WITH OBSERVATIONS}
To date more than a dozen KBBs have well determined orbits
\citep[e.g.][]{N07}. Unfortunately due to projection effects, the prograde and
retrograde orbital solutions of the KBBs are nearly degenerate. This
degeneracy can usually only be broken after several years once the viewing
angle of the KBBs has changed sufficiently. Very recently, after the
submission of our original manuscript, two groups reported unique orbital
solutions for KBBs Typhon-Echidna \citep{GN08} and $2001~QW_{322}$
\citep{P08}. \citet{GN08} find a prograde orbit for Typhon-Echidna and
\citet{P08} report a retrograde orbit for $2001~QW_{322}$. $2001~QW_{322}$ has
such a large binary separation that, even in the current Kuiper belt, it
experiences significant dynamical interactions with other large KBOs. It is
early to draw conclusions for the whole binary population, but if a comparable
number of retrograde and prograde binaries is found it would imply that KBBs
formed from a dispersion dominated KBO disk, which would also be consistent
with observed binary inclinations. Dispersion dominated KBO velocities would
imply that the value of $\Sigma/\sigma$ was larger during binary formation
than what we used in equation \ref{e3}. However, the velocity dispersion
during binary formation cannot have exceeded $v_H$ significantly since the
binary formation timescales would otherwise become excessively long
\citep{N07,SR07}.

\section{DISCUSSION AND CONCLUSIONS}
The relative abundance of prograde to retrograde orbits enables us to
differentiate between various proposed binary formation scenarios
observationally.

We predict that the vast majority ($\gtrsim 97$ \%) of comparable mass ratio
binaries will have retrograde orbits if KBO velocities of $\lesssim 0.1 v_H$
prevailed during their formation. This dominance of retrograde over prograde
binary orbits is due to the fact that for sub-Hill velocities binaries form
primarily via the $L^2s$ mechanism instead of $L^3$ mechanism. Since dynamical
friction only gradually increases the modified Jacobi constant, all binaries
that form via the $L^2s$ mechanism, or any other mechanism that dissipates
little energy over an orbital timescale, will start with modified Jacobi
constants close to that of unbound KBOs. Only stable retrograde orbits exist
for modified Jacobi constants similar to that of KBOs with initially circular
orbits around the Sun that penetrate inside the Hill sphere. Therefore, KBBs
have retrograde orbits provided that they form from dynamically cold KBOs via
the $L^2s$ mechanism.

As the KBO velocities approach $v_H$ the preference of retrograde orbits
decreases. Further, we predict a comparable number of
prograde and retrograde binaries form for super-Hill KBO velocities. This is
because only the $L^3$ mechanism can form tight binaries that tend to
survive if super Hill velocities prevail \citep{SR07}. The fact that
retrograde orbits are stable for larger semi-major axes is no longer of
importance since only tight binaries tend to survive. This therefore leads to
the formation of a roughly equal number of prograde and retrograde binaries
for super-Hill KBO velocities.

\begin{figure} [htp]
\plotone{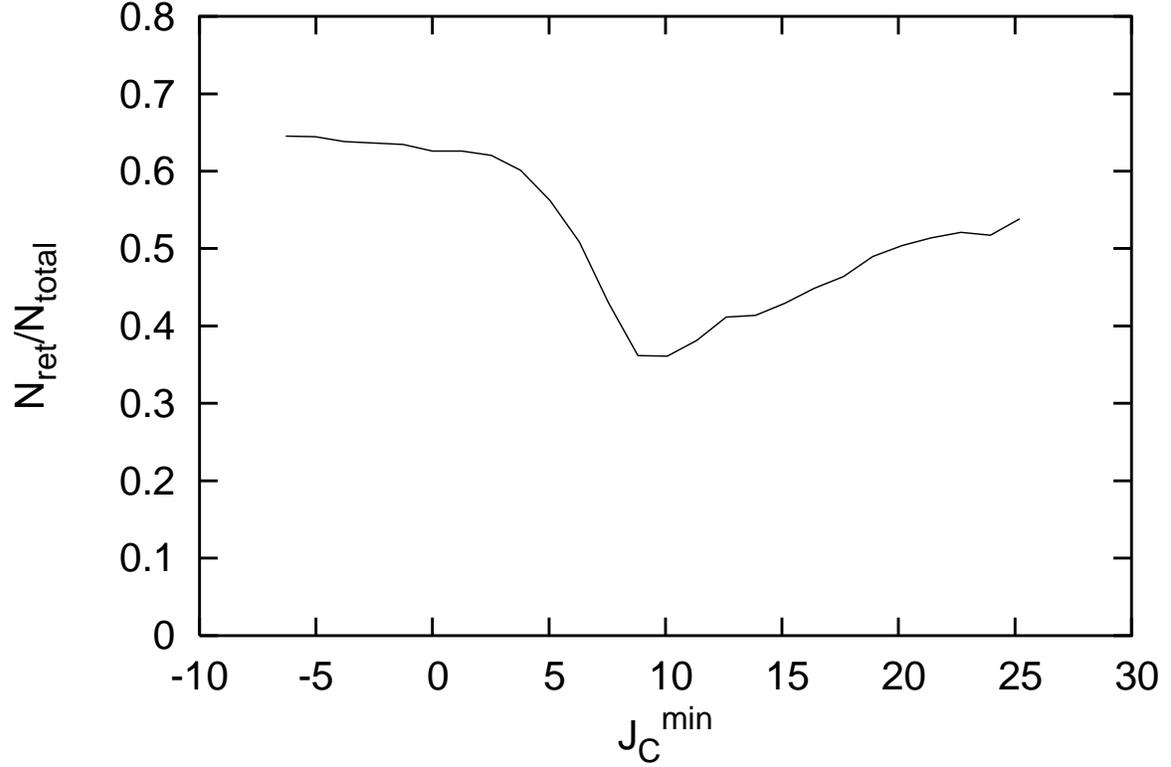} 
\caption{The ratio of retrograde binaries, $N_{ret}$, with a modified Jacobi
constant of $J_C^{min}$ or larger to the total number of binaries, $N_{total}$,
that formed via the $L^3$ mechanism for $v \ll v_H$. For small $J_C^{min}$,
i.e. when all binaries are included, about $2/3$ have retrograde orbits. More
retrograde than prograde binaries form because retrograde binary orbits are
stable further out in the Hill sphere than prograde ones. As $J_C^{min}$
increases the fraction of retrograde binaries decreases reaching a minimum of
about $1/3$ for $J_C^{min} \sim9$. This may be due to the Keplerian sheer
which increases the duration of a prograde encounter between unbound KBOs
compared to a retrograde encounter. The fraction of prograde and retrograde
binaries becomes comparable for $J_C^{min} \gg 9$ because for such binaries
neither the Keplerian sheer nor the increased stability of retrograde orbits
are important.}
\label{fig4} 
\end{figure}

The analysis presented here has also implications for some of the other
proposed binary formation scenarios. \citet{W02} suggested that KBBs form by a
collision among two KBOs inside the Hill sphere of a third. Although the $L^3$
mechanism dominates over such a collisional binary formation scenario we
briefly discuss our predictions for this collisional binary formation
mechanism. For sub-Hill velocities more retrograde than prograde binaries form
because retrograde binary orbits are stable further out in the Hill sphere
than prograde ones (i.e. the phase space for forming retrograde binaries is
larger than that for prograde binaries). For super-Hill velocities a
comparable number of prograde and retrograde binaries form because the fact
that retrograde orbits are stable for larger semi-major axes is no longer of
importance since only tight binaries are saved from break up. In the formation
scenario of \citet{A05} the existence of long lived transient binaries that
spend a long time in their mutual Hill sphere, near a periodic orbit, is
responsible for the creation of KBBs. \citet{A07} find an excess of prograde
over retrograde binaries and suggest that this is a signature of their binary
formation process. Our work indicates that an excess of prograde over
retrograde binaries might simply be the result of the velocity regime (i.e.$v
\sim v_H$) in which the binaries form (see Figure \ref{fig4}).

All of the above predictions rely on the assumption that subsequent
orbital evolution due to dynamical friction and dynamical stirring of the
Kuiper belt did not alter the sense of the binary orbit. The Kuiper Belt has
undergone a phase of dynamical excitation which probably modified the orbital
properties of KBBs. A detailed study on how dynamical stirring of the Kuiper
Belt and dynamical friction affects binary inclinations would be very
worthwhile to determine whether they were able to reverse the binary orbit
from prograde to retrograde rotation.

\acknowledgments 
Some of the numerical calculations presented here were performed
on Caltech's Division of Geological and Planetary Sciences Dell
cluster. R. S. is an Alfred P. Sloan Fellow and a Packard Fellow.

\clearpage 

\end{document}